\documentclass[10pt]{llncs}

\usepackage{amsmath}
\usepackage{amssymb}
\usepackage{longtable}

\usepackage{setspace}
\usepackage{cite}

\usepackage{graphicx}
\usepackage{subfigure}
\usepackage{epstopdf} % auto-convert eps to pdf
\usepackage{dcolumn} % Align table columns on decimal point

\usepackage{color} 
\makeatletter
\renewcommand{\@biblabel}[1]{\quad#1.}
\makeatother

\begin{document}

\begin{flushleft}
{\Large
\textbf{Scaling of foreign attractiveness for countries and states}
}
\\
\vspace{15pt}

Iva Bojic$^{1, 2}$, 
Alexander Belyi$^{2,3}$,
Carlo Ratti$^{1}$,
Stanislav Sobolevsky$^{1,4}$
\vspace{10pt}
\\
${ }^{1}$Senseable City Laboratory, Massachusetts Institute of Technology, 77 Massachusetts Avenue, Cambridge, MA 02139, USA
\\
\vspace{10pt}
${ }^{2}$SENSEable City Laboratory, SMART Centre, 1 Create Way, Singapore
\\
\vspace{10pt}
${ }^{3}$Belarusian State University, 4 Nezavisimosti Avenue, Minsk, Belarus
\\
\vspace{10pt}
${ }^{4}$Center for Urban Science and Progress, New York University, 1 MetroTech Center, Brooklyn, NY 11201, USA 
\vspace{10pt}
\\
$\ast$ E-mail: ivabojic@mit.edu, alex.bely@smart.mit.edu, ratti@mit.edu, sobolevsky@nyu.edu
\end{flushleft}

\section*{Abstract}                                                                                                                                                                                                                                         
People's behavior on online social networks, which store geo-tagged information showing where people were or are at the moment, can provide information about their offline life as well. In this paper we present one possible research direction that can be taken using Flickr dataset of publicly available geo-tagged media objects (e.g., photographs, videos). Namely, our focus is on investigating attractiveness of countries or smaller large-scale composite regions (e.g., US states) for foreign visitors where attractiveness is defined as the absolute number of media objects taken in a certain state or country by its foreign visitors compared to its population size. We also consider it together with attractiveness of the destination for the international migration, measured through publicly available dataset provided by United Nations. By having those two datasets, we are able to look at attractiveness from two different perspectives: short-term and long-term one. As our previous study showed that city attractiveness for Spanish cities follows a superlinear trend, here we want to see if the same law is also applicable to country/state (i.e., composite regions) attractiveness. Finally, we provide one possible explanation for the obtained results.

Flickr; scaling laws; media objects; big data

\section{Introduction}
\label{sec:introduction}

% human mobility thing
It is true that advancement of transport systems in the last 100 years provided people with more mobility options, but it was only a little bit more than 5~years ago, with the emergence of social network platforms, shared economy and multiple mobile devices, that digital records on human mobility became public in a real-time. Namely, in a society relying on shared economy traveling cost per capita are reduced which resulted in a higher number of people who actually move and while doing that they use their smartphones equipped with Global Position Systems (GPS) enabling them to capture their geo-location and publicly share it on their social networks. Although it is still not clear how \textit{living online} will affect us in future, we can be sure for one thing -- the huge digital data produced by people all around the world represents almost an inexhaustible resource that can be used in different scientific studies.

Different types of such data are now being increasingly utilized for estimating human mobility, including cell phone data \cite{gonzalez08, calabrese2011estimating, kung2014exploring, hoteit2014estimating}, vehicle GPS traces \cite{santi2014quantifying, kang2013exploring} or credit card transactions \cite{sobolevsky2015cities, lenormand2015influence}. However, so far geo-tagged social media was one of a few sources allowing to approach human mobility on a global scale \cite{hawelka2014geo, Paldino2015Urban, belyi2016global}. Applications of those studies include diverse examples such as urban transportation \cite{santi2014quantifying}, determining economical potential of the cities \cite{sobolevsky2015predicting}, land use classification methods \cite{grauwin2015towards, pei2014new}, or regional delineation \cite{ratti2010redrawing, amini2014impact}.

% story about cities
Geo-location part of information on human mobility allows scientific studies to focus on specific areas, such as on cities which are important as today more than 50\% of the world population lives in them. The trend of the urban population growth is expected to be continued in future as well and it is estimated that in the next 30 to 40 years the number of city dwellers might nearly double \cite{un2014world}. However, cities are not only studied because of the number of people living there, but also because of their long and rich history. Many cities around the world started their development back in medieval or classical times and during that long span of time experienced many demographic, economic, political and spatial transformations \cite{hall1998cities}. Nevertheless, cities are not only exceptional and unique, but they also do share certain common properties as shown in scientific studies across different disciplines (e.g., economics, engineering, complex systems). These effects are known as agglomeration or scaling relation between macroscopic properties of cities (e.g., employment rate) and its size \cite{bettencourt2013origins}. 

% other studies on scalability
Although over the years urban scaling theory shed lights on superlinear behavior of socioeconomic (e.g., intensity of interactions \cite{schlapfer2012scaling}, creativity \cite{bettencourt2010urbscaling}, economic efficiency \cite{bettencourt2013origins}) and sublinear scaling of urban infrastructure \cite{batty2008size}, it is still not clear if this behavior could be extended to other scales, such as provinces or the whole states and countries. In this study we focus on investigating scaling laws of attractiveness, defined as the number of digital media objects created by foreign visitors from Flickr dataset or number of foreign citizens/foreign born population living in different countries across the world captured in dataset created by United Nations. Our previous studies presented results on city attractiveness in Spain, showing strong superlinearly dependence (with scaling exponent around 1.5) \cite{Sobolevsky2015Scaling, sobolevsky2014mining, bojic2015sublinear} and here we extend our analysis to other spatial scales as well. 

% the short idea of our paper
In this paper we discuss possible reasons why and at which level attractiveness starts to scale sublinearly. Namely, city attractiveness scales superlinearly in contrast to country level that shows a sublinear behavior. In order to find a turning point between cities and countries, in our analysis we included the US states as an intermediate level and showed that on that level the scaling exponents indeed lay between cities' and countries' exponents. The rest of the paper is organized as follows. Section \ref{data_sets} describes two datasets used to check country/state scaling laws. Furthermore, Section \ref{attract} shows results for country attractiveness and scaling of attractiveness of the US states. Finally, Section \ref{conclusion} discusses our results and provides final conclusion marks.

\section{Datasets}
\label{data_sets}

In this study we use publicly available Yahoo Flickr Creative Commons dataset of the largest public multimedia collection that has ever been released and created as a part of the Yahoo Webscope program\footnote{Available at http://webscope.sandbox.yahoo.com.}. It contains 100 million media objects that were created between 2004 and 2014 where each media object in the dataset is represented by its metadata: object identifier, user identifier, time stamp when it was taken, location (i.e., latitude and longitudinal coordinates) where it was taken (if available), and CC license it was published under \cite{thomee2016YFCC100M}. Additionally, the metadata for some objects also contains object title, user tags, machine tags and description, as well as a direct link for downloading the content. In addition to the amount of media objects and supporting metadata for each object, the main advantage of this dataset compared to the other similar ones is that all media objects are licensed under one of the Creative Commons copyright licenses, and as such can be used for benchmarking purposes as long as media objects are credited for the original creation.

Once when downloaded the raw dataset, we pruned the data by first omitting records that were not geo-tagged (i.e., more than 50\%) and then by omitting those that came with the wrong date format (less than 0.01\%). Since the focus of the paper is to investigate how attractiveness scales on a country level, we had to perform reverse geo-coding so that every geo-location (i.e., latitude and longitude coordinates) is translated into an area we observed (e.g., state or country). We used shapefiles that define boundaries of world countries\footnote{Shapefiles are available at http://thematicmapping.org/downloads.} and the US states\footnote{Shapefiles are available at http://gadm.org/country.}. After both pruning and reverse geo-coding we were left with 40 million objects created in $238$ countries around the world. In addition to Flickr dataset, as another source for verification of a sublinear country scaling law, we are using official migration statistics provided by United Nations\footnote{Data can be download at: http://www.un.org/en/development/desa/population/ migration/data/estimates2/estimatesorigin.shtml.}. This statistic is provided in a form of an origin/destination (OD) matrix which represents an estimate of how many foreign citizens had lived in each country in July of 2010.

\section{Attractiveness scaling laws}
\label{attract}

As defined in our previous paper \cite{Sobolevsky2015Scaling} foreign attractiveness for Flickr dataset is quantified as the number of media objects created by people visiting the considered area (e.g., country) from other countries. Similarly, attractiveness for migration dataset is represented with the number of the current country residents who moved to that country from abroad. Since Flickr dataset does not include information about where people live, we used the following methodology of inferring the user's home country~-- we say that a particular user lives in a particular country if there he/she created the maximum number of media objects and spend the maximal number of days compared to all other countries across the world (if maximal activity and activity period give inconsistent results, we consider that home country is undefined). More detailed discussion on the importance of choosing the right home definition method can be found in \cite{bojic2015choosing}. Once when finished with inferring the home location, we took the activity only of those users for whom we detected where they lived (we were left with $72.4\%$ of original objects) and calculated aggregated attractiveness as the number of media objects taken by foreign users ($15.2\%$ of all objects). Aggregated attractiveness means that we took the total aggregated user activity over the entire timeframe of the data availability without taking into consideration time when it as made.

\begin{figure*}[h!]
\centering
	\includegraphics[width=1\textwidth]{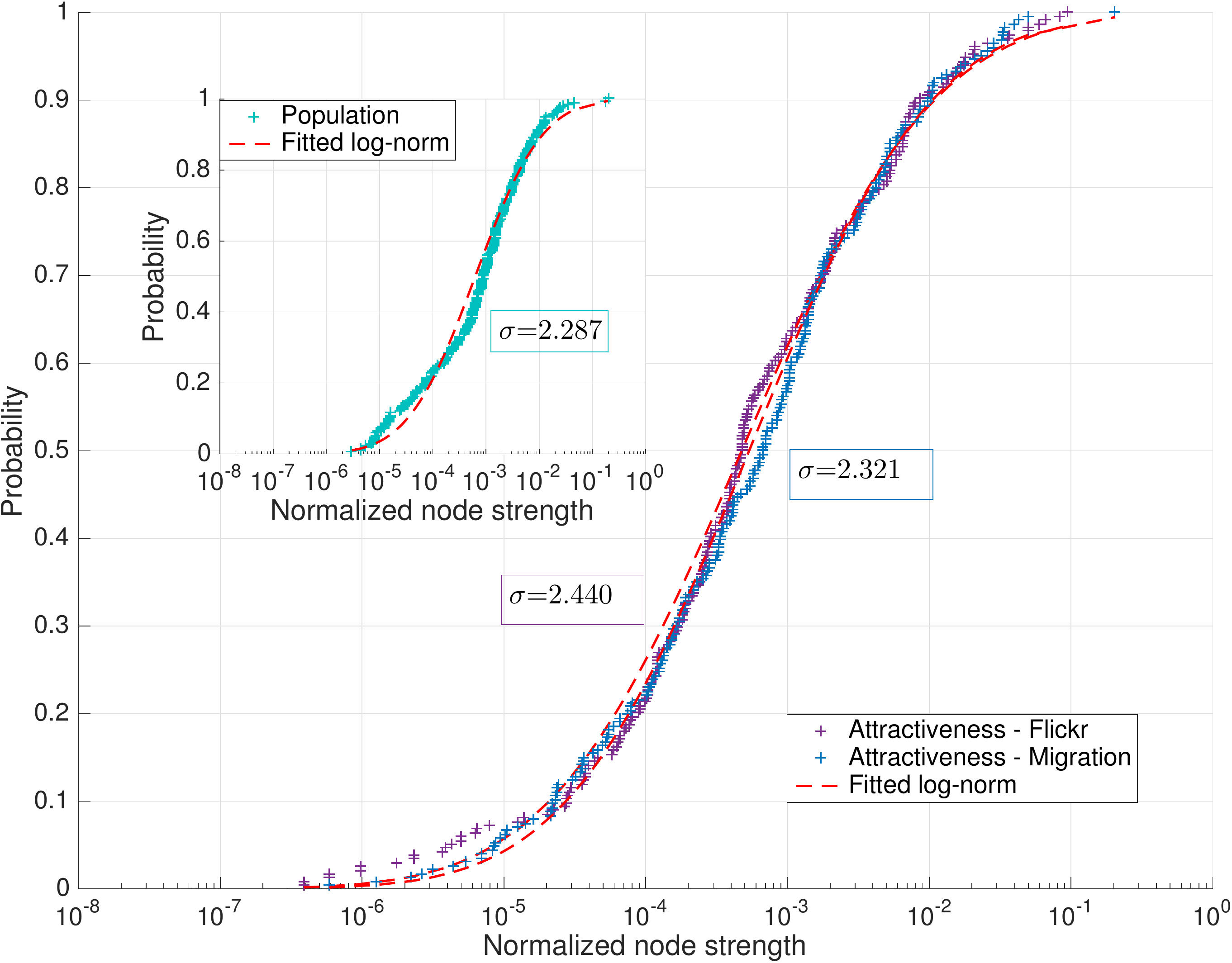}
	\caption{\label{fig:distrib}Distribution of countries populations and attractiveness based on Flickr and migration data. All three variables are distributed very similarly with distribution well approximated by log-normal distribution with standard deviation~$\sigma$ around~$2.3-2.4$.}
\end{figure*}

Flickr dataset and dataset provided by United Nations (i.e., migration dataset), which are described in the previous section, reveal different aspects of country attractiveness~--- short-term and long-term one. Namely, rather than showing mostly a touristic behavior, a migration dataset counts in for people who decided to temporarily or more permanently move to another country. Even before trying to find scaling law patterns for both datasets, in Fig. \ref{fig:distrib} we show distribution of country attractiveness as well as its population. Both types of attractiveness are distributed similarly in shape and follow the same log-normal patterns as country population distribution does~--- although variance for them is a slightly different: a higher one for attractiveness. Now let us consider how attractiveness of each specific country is actually related to its population.

\begin{figure*}[h!]
	\centering
	\subfigure[Flickr]{
		\label{subfig:scaling_countries_flickr}
		\includegraphics[width=.475\textwidth]{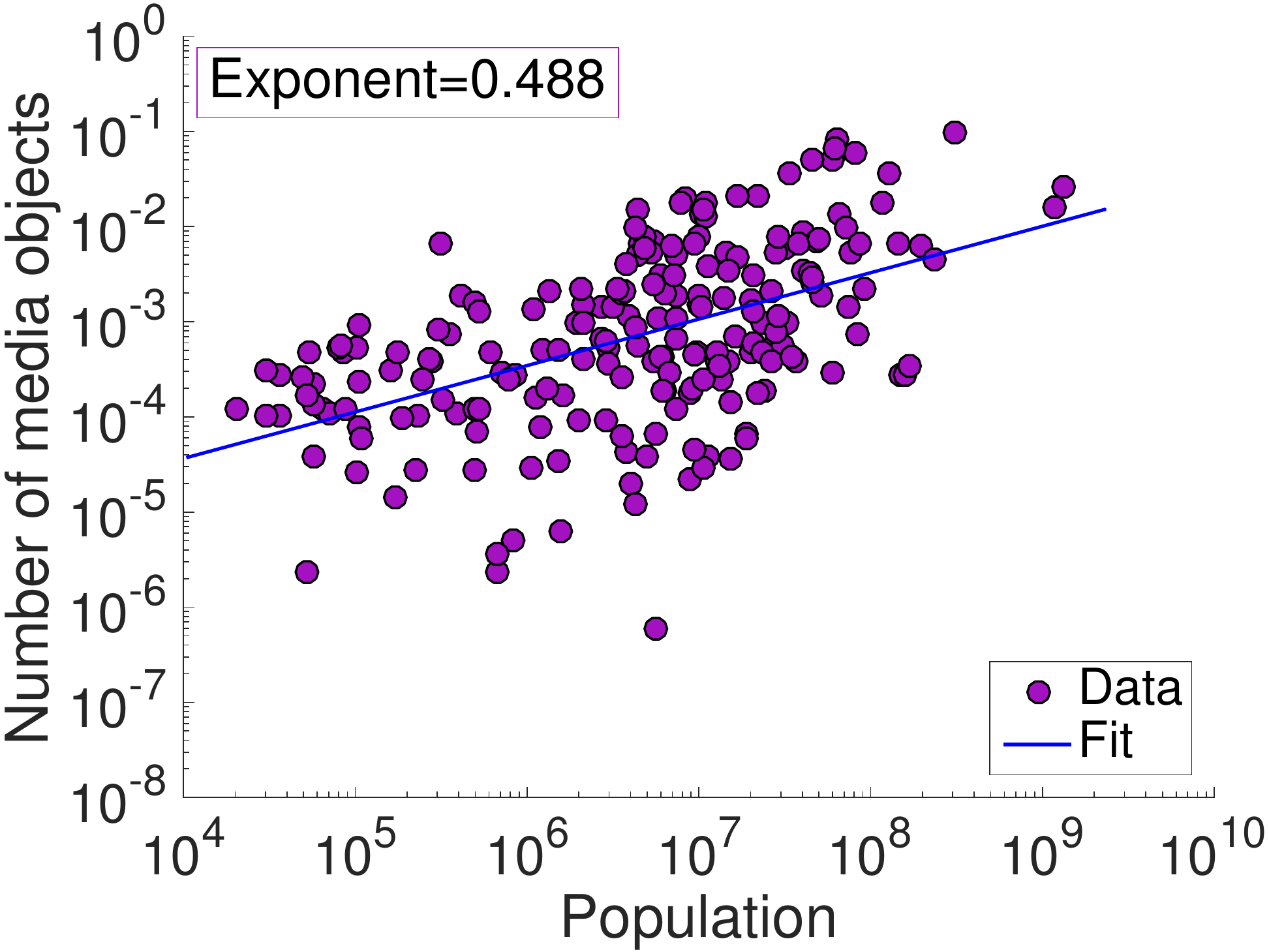}
	}
	\subfigure[Migration]{
		\label{subfig:scaling_countries_migration}
		\includegraphics[width=.475\textwidth]{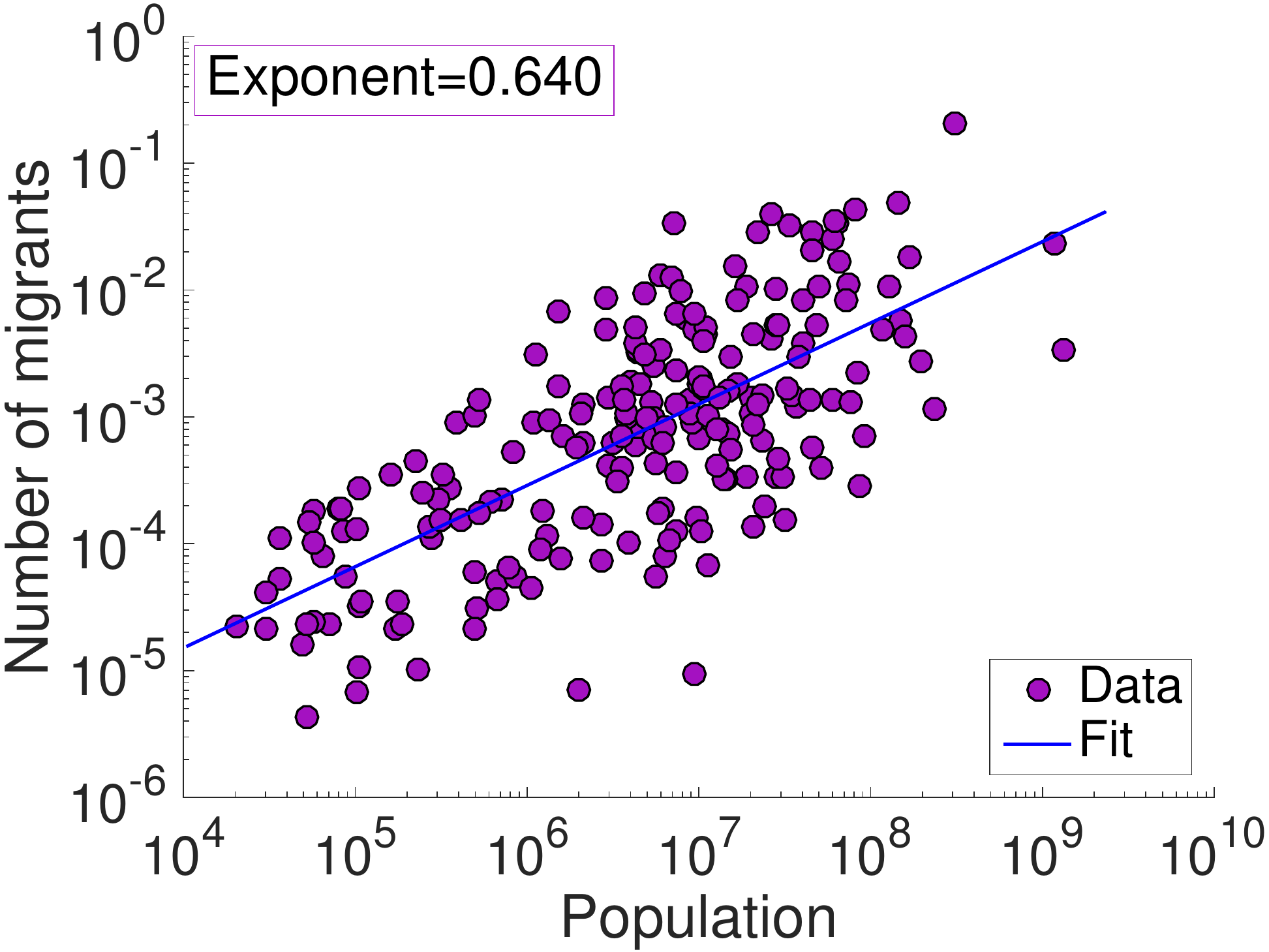}
	}
	\caption{\label{fig:scaling_countries}
	Results of country attractiveness scaling calculated from Flickr datasets are presented in Fig.\,\ref{subfig:scaling_countries_flickr} and for migration dataset in Fig.\,\ref{subfig:scaling_countries_migration}. There are $238$ countries in both datasets and both plots show very strong sublinear scaling with exponents equal to~$0.488$ and~$0.640$ ($95\%$ confidence intervals are~$[0.377~0.599]$ and~$[0.551~0.730]$~-- both considerably below 1~-- validating the robustness of sublinear scaling, while $R^2=0.27$ and $R^2=0.49$ respectively hinting for the high variation in the specific country attractiveness around the trend).}
	\vspace{-10pt}
\end{figure*}

The results of fitting a power-law dependence $A\sim a p^{\beta}$ to the total number of media objects denoting aggregated attractiveness~$A$ and population~$p$ are shown in Fig.\,\ref{fig:scaling_countries}. Fitting process is performed on a log-log scale where it becomes a simple linear regression problem $log(A)\sim log(a)+\beta\cdot log(p)$. As explained in \cite{bettencourt2007growth} depending on the value of parameter $\beta$ scaling can be: sublinear ($\beta < 1$), linear ($\beta = 1$) or superlinear ($\beta > 1$). Therefore, unlike the results presented in \cite{Sobolevsky2015Scaling} showing that urban attractiveness demonstrates a superlinear behavior (just like the other socioeconomic urban properties), the country attractiveness is scaling sublinearly. This finding is true for both datasets representing short and long term attractiveness and the confidence intervals for the exponent validate the robustness of a conclusion about sublinear scaling.

\begin{figure*}[h!]
	\centering
	\includegraphics[width=1\textwidth]{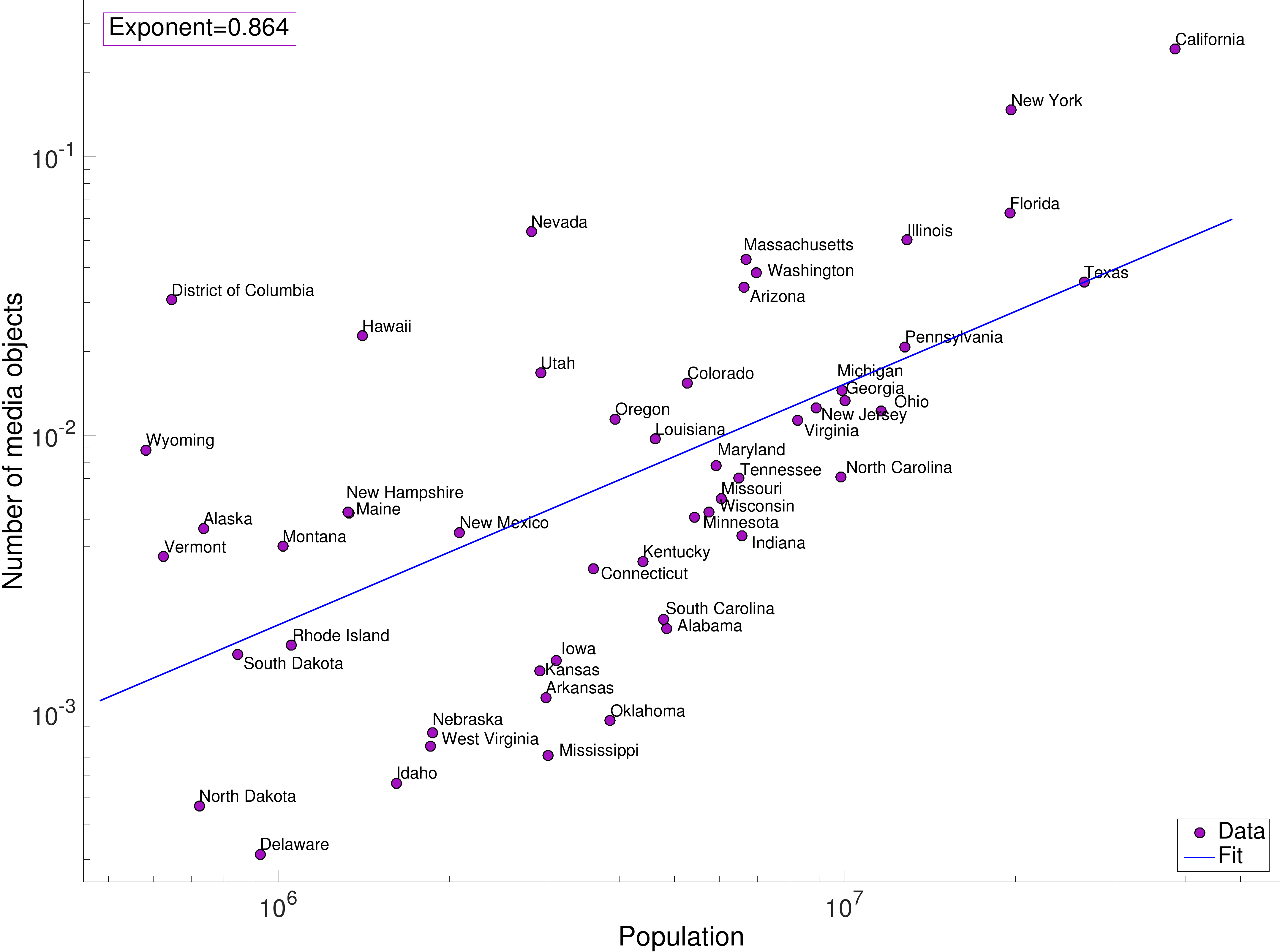}
	\caption{\label{fig:state_scaling}The sublinear scaling of US states attractiveness with population size for 50 US states and Washington~D.\,C. with exponent estimate of~$0.864$ ($95\%$ confidence interval is~$[0.530~1.198]$, $R^2=0.36$).}
\end{figure*}

\begin{figure*}[h!]
	\centering
	\includegraphics[width=1\textwidth]{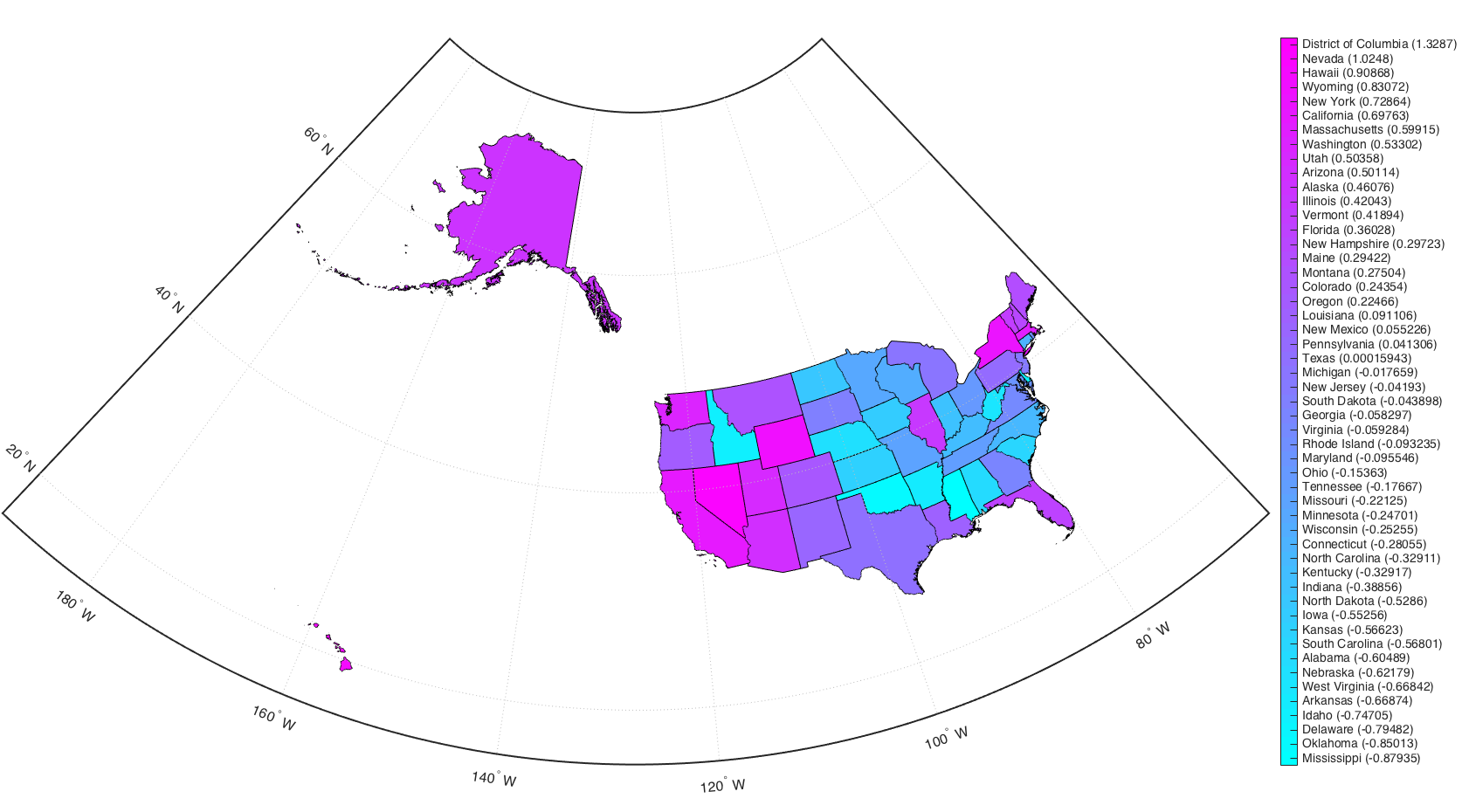}
	\caption{\label{fig:state_performance}Relative attractiveness of the US states measured as difference between logarithms of observed and expected number of Flickr media objects.}
\end{figure*}

In addition to our finding that country attractiveness does not follow the same scaling pattern as an urban one, Fig.\,\ref{fig:state_scaling} presents an interesting finding that attractiveness for the areas of an intermediate spatial level~-- the US states~-- falls between scaling exponents for cities and countries. This could suggest that there are factors other than just population which play big roles in attractiveness laws. Moreover, the high variation of the specific attractiveness values around the average trend points that it might be interesting to consider which states are over-performing and which ones are under-performing (see Fig.\,\ref{fig:state_performance}). Such an approach allows to come up with a more objective ranking of state attractiveness free from the effect of scale, i.e., comparing states not only by their absolute attractiveness values, which are highly dependent on the state size, but comparing them with the baseline first and using the over-/under-performance measures that quantified a log-scale residual from the scaling trend as a scale-free performance characteristic.
 
The top three over-performing areas, also being the strongest upper outliers in Fig.\,\ref{fig:state_scaling} are Washington~D.\,C., Nevada and Hawaii with Wyoming, New York and California going right after. The very likely reason behind the over-performance of Washington~D.\,C., Nevada and Hawaii could be their relatively small population size compared to a very high number of tourists visiting them. Also New York and California are certainly the major destinations too, while it is not so obvious and might be an interesting study to understand why Wyoming happens to be among the leading countries as well. On the other hand, the most under-performing states in terms of their foreign attractiveness quantified through Flickr data are Delaware, Oklahoma and Mississippi. In general, the patterns we observed are pretty consistent with those which one would expect to see from the touristic context of the areas considered.

\begin{figure*}[h!]
	\centering
	\subfigure[]{
		\label{subfig:num_of_cities}
		\includegraphics[width=.475\textwidth]{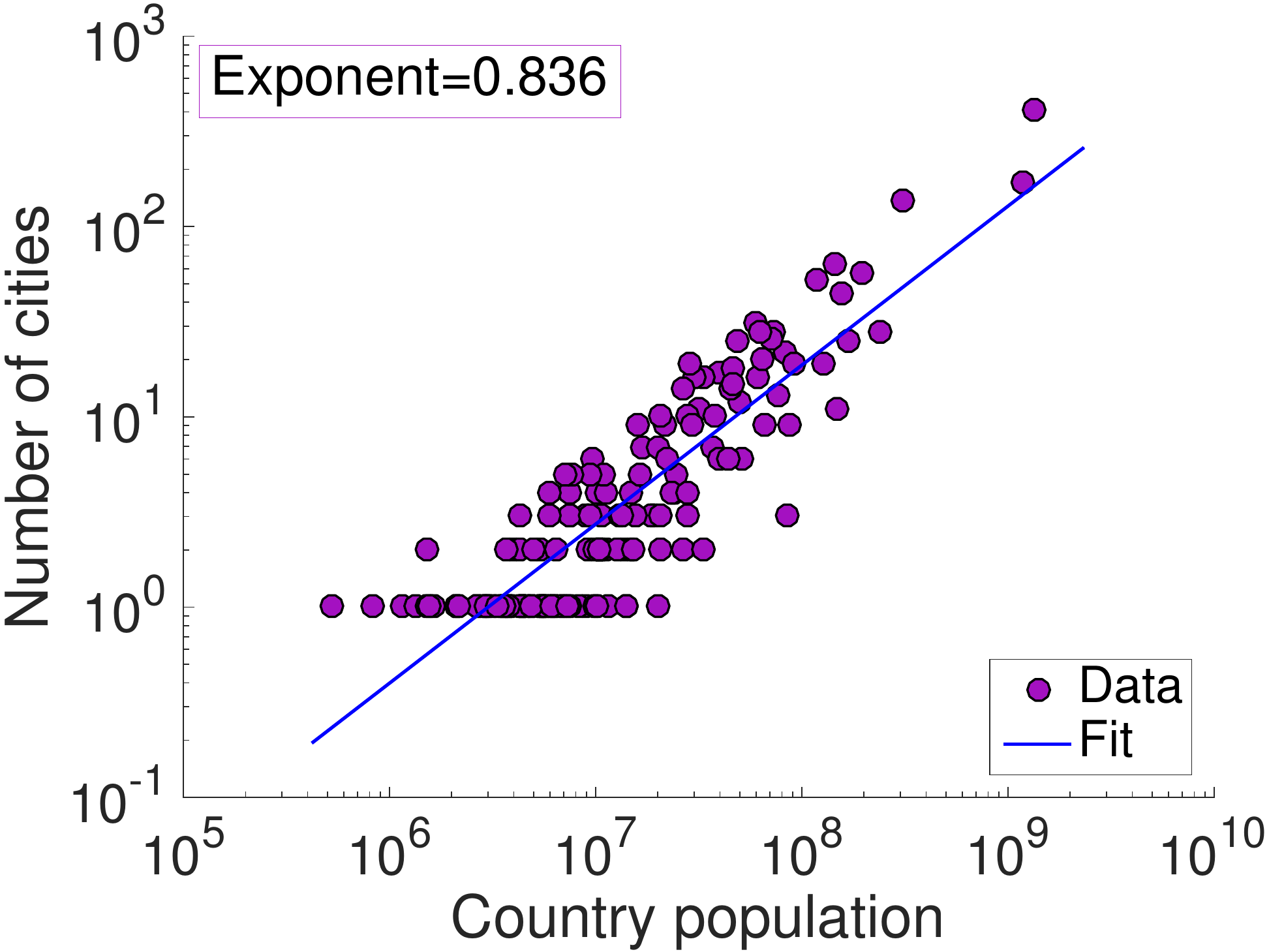}
	}
	\subfigure[]{
		\label{subfig:capital_population}
		\includegraphics[width=.475\textwidth]{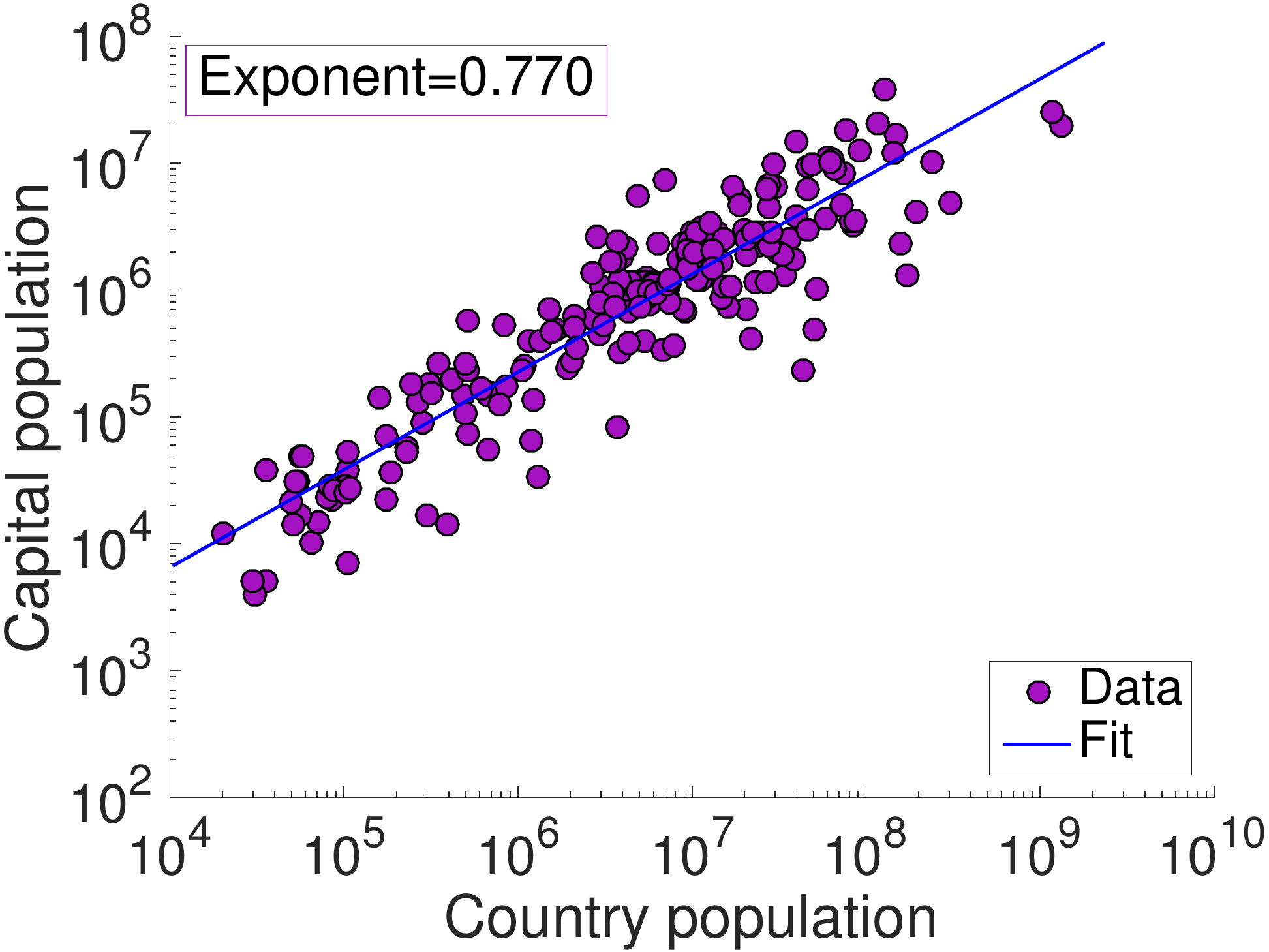}
	}
	\caption{\label{fig:city_scaling}Scaling of number of cities Fig.\,\ref{subfig:num_of_cities} and population of capital Fig.\,\ref{subfig:capital_population} with country population. Both plots show very significantly sublinear scaling with exponents equal to~$0.836$ and~$0.770$ ($95\%$ confidence intervals are~$[0.762~0.910]$ and~$[0.722~0.818]$, while high $R^2$ values of $0.77-0.83$ show that the trends provide a decent insight on the structure for most of the countries) correspondingly.}
\end{figure*}

In order to provide one possible explanation about differences between country and city scaling behaviors, we zoomed in into a composite structure of countries around the world, which makes them different from the cities. Namely, Fig.\,\ref{fig:city_scaling} presents a sublinear scaling exponent for the number of cities and sizes of capital cities. We used data from United Nations website\footnote{Data can be download at: http://esa.un.org/unpd/wup/CD-ROM} and because only cities with population more than 300,000 were considered, there are a lot of countries with just one or two of such cities. These findings are to a certain extent in line with the sublinear scaling law of urban infrastructure presented in \cite{batty2008size} if an urban decomposition of the country is seen as its own infrastructure in a sense.

Results presented in this paper together with previous observations \cite{Sobolevsky2015Scaling} show that foreign tourists are mostly concentrated on the major destinations across the country, giving an idea for the possible reasons of the countries sublinear attractiveness scaling phenomena. Being increasingly heterogeneous composite hierarchical structures, larger countries provide relatively smaller numbers and size of top destinations of major interest for foreign visitors, compared to what one would expect from a simple proportionality relation to the country population, whereas the majority of country population is being spread across the areas that foreign visitors are not that familiar with or do not find appealing enough to pay their visit to.

\section{Discussion and conclusions}
\label{conclusion}

The overall goal of our study is to provide data necessary for enhancing touristic planning by taking into a consideration user individual preferences using information from an unconventional big data source: dataset of geo-tagged media objects that people take at places they visit. As the number of people publicly sharing their media objects increases on a daily basis, new almost real-time insights on human behavior while traveling are becoming available. This means that rather than having the statistic once a month or on a yearly basis, nowadays urban planners and touristic workers are empowered to make decisions almost simultaneously with the emergence of new events reflected through social media. Moreover, results of our study can be used not only to compare how popular different touristic places are, but also to predict how popular they should be taking into an account their characteristics, such as population size. And the latter can even help city officials and touristic workers to detect problems (if there are any) and try to find solutions for them.

Many related studies looked into demographic, economic and political parameters and how they scaled with city population. Conclusions of those studies were very consistent showing that: (1) parameters closely related to a city size (such as number of available jobs) scale linearly with city population, (2) socioeconomic properties (such as urban attractiveness) show superlinear behavior, while (3) parameters of urban infrastructure on the other hand are scaling sublinearly. However, not many studies investigated how a certain parameter scaled across different spatial levels (e.g., city and country) which is exactly the focus of our study on scaling behavior of the attractiveness of composite areas, such as countries or states.

As shown earlier, attractiveness of cities in Spain, defined as the number of media objects taken in that city and stored in Flickr dataset, scaled superlinearly with the city size. Moreover, this was true even for different definitions of cities. On the other hand, as shown in this paper, exponents for country attractiveness are less than 1 for both Flickr dataset and dataset of human migration provided by United Nations showing a consistent pattern across different datasets. In that context a rather expected result was that a scaling exponent for the foreign attractiveness of the US states, which are smaller, but still composite areas, would fell somewhere in between values of exponents for urban and country attractiveness, but still tend to be slightly sublinear similar to the case of countries.

Finally, in that context, our study also analyzed the possible reasons for this difference between country and city attractiveness. Unlike cities, the countries are more heterogeneous structures, including not only highly attractive bigger cities, but also a large number of other places that foreigners might not consider as very attractive areas. Moreover, when looking at some quantitative properties of country structures, such as the number of their cities or the size of their biggest cities, we can also observe a sublinear scaling pattern, which might be the reason why the total country attractiveness, being mainly defined by its bigger cities, also scales sublinearly. The US states being somewhere on an intermediate level between country and city scales, are also composite structures, but not that large-scale and heterogeneous, and that is why their attractiveness scaling exponent is still less than 1, but considerably higher than in case of countries.

\section*{Acknowledgments}
The research is supported by the National Research Foundation, Prime Minister's Office, Singapore, under its CREATE programme, Singapore-MIT Alliance for Research and Technology (SMART) Future Urban Mobility (FM) IRG. The authors would like to thank MIT SMART Program, Center for Complex Engineering Systems at KACST and MIT, Accenture, Air Liquide, BBVA, The Coca Cola Company, Emirates Integrated Telecommunications Company, The ENEL foundation, Ericsson, Expo 2015, Ferrovial, Liberty Mutual, The Regional Municipality of Wood Buffalo, Volkswagen Electronics Research Lab, UBER, and all the members of the MIT Senseable City Lab Consortium for supporting the research. Finally, this work was partially supported by research project "Managing Trust and Coordinating Interactions in Smart Networks of People, Machines and Organizations", funded by the Croatian Science Foundation under the project UIP-11-2013-8813. 

\bibliography{literature}

\end{document}